\documentclass[a4paper,11pt]{article}
\usepackage{pos}
\usepackage{float}
\usepackage{graphicx}
\usepackage{subcaption}

\newcommand{\preprint}{{\scriptsize \texttt{CERN-TH-2026-232, WUB/26-05}}}

\title{Improving hadron creation operators for charmonium, glueballs and baryons}

\author*[a,b,d]{Juan Andrés Urrea-Niño}
\author[f]{John Bulava}
\author[c]{Nikolai Husung}
\author[d]{Francesco Knechtli}
\author[d]{Tomasz Korzec}
\author[a,b]{Michael Peardon}
\author[e]{Fernando Romero-López}
\author[e]{Miguel Salg}

\affiliation[a]{School of Mathematics, Trinity College Dublin, College Green, Dublin 2, Dublin, Ireland}

\affiliation[b]{Hamilton Mathematics Institute, Trinity College Dublin, College Green, Dublin 2, Ireland}

\affiliation[c]{Theoretical Physics Department, CERN, 1211 Geneva 23, Switzerland}

\affiliation[d]{Department of Physics, Bergische Universität Wuppertal
Gaußstraße 20, 42119 Wuppertal, Germany}

\affiliation[e]{Albert Einstein Center, Institute for Theoretical Physics, University of Bern, 3012 Bern, Switzerland}

\affiliation[f]{Institut für Theoretische Physik II, Ruhr-Universität Bochum, D-44780 Bochum, Germany}

\emailAdd{urrea.nino@tcd.ie}
\emailAdd{john.bulava@rub.de}
\emailAdd{nikolai.husung@cern.ch}
\emailAdd{knechtli@uni-wuppertal.de}
\emailAdd{korzec@uni-wuppertal.de}
\emailAdd{mjp@maths.tcd.ie}
\emailAdd{fernando.romero-lopez@unibe.ch}
\emailAdd{miguel.salg@unibe.ch}

\abstract{Constructing hadron creation operators which properly sample the finite-volume energy spectrum is a fundamental step in lattice QCD spectroscopy and scattering calculations. Operators enabling access to spectrum information from correlation functions at early temporal separations are extremely advantageous, particularly when the signal-to-noise problem is severe. We present improved operator constructions for charmonium, glueball and baryon spectroscopy which exhibit these advantages. Novelties of our work include the extension of the distillation profiles framework to baryon and multi-meson calculations, as well as an efficient implementation of glueball-like operators which retain angular momentum information from the continuum.}

\FullConference{The 43rd International Symposium on Lattice Field Theory (Lattice 2026)\\
July 26 to August 1, 2026\\
University of Maryland, College Park, USA\\}

\begin{document}
\begin{flushright}
\preprint
\end{flushright}
\maketitle

\section{Introduction}
Computing Euclidean two-point temporal correlation functions is fundamental for hadron spectroscopy as their asymptotic behavior is dictated by the energies of the states of interest. Two issues often arise in these calculations: how to properly choose the hadron creation operators used so that physical information of states of interest is accessible at short time separations (operator problem), and the statistical signal-to-noise problem by which the error relative to the signal grows exponentially with the time separation (noise problem). These two issues are related in a window problem; at short time separations the statistical error is small but commonly used operators lead to significant excited-state contamination, while at large time separations the low-energy states of interests dominate the signal but the large errors make it inaccessible. These problems affect correlation functions coming from all types of operators, particularly those containing so-called disconnected contributions such as those for flavor-singlet meson operators or purely gluonic ones commonly used for glueball-related calculations. In this work we tackle the operator problem for three different systems of current interest: nucleons, glueballs and charmonium. For nucleons, we introduce optimal distillation profiles and significantly reduce excited-state contamination at different values of spatial momentum. For glueballs, we present an efficient implementation of gluonic operators which retain angular momentum information from the continuum and improve over commonly-used spatial loops. For charmonium, we extend the framework of optimal distillation profiles to the 2-meson case using highly-improved 1-meson operators combining optimal profiles with the already well-established derivative-based constructions. In all cases we observe significant improvement compared to currently used operators.

\section{Results}
\subsection{Improved nucleon operators}
\begin{figure}
    \centering
    \includegraphics[width=0.9\textwidth]{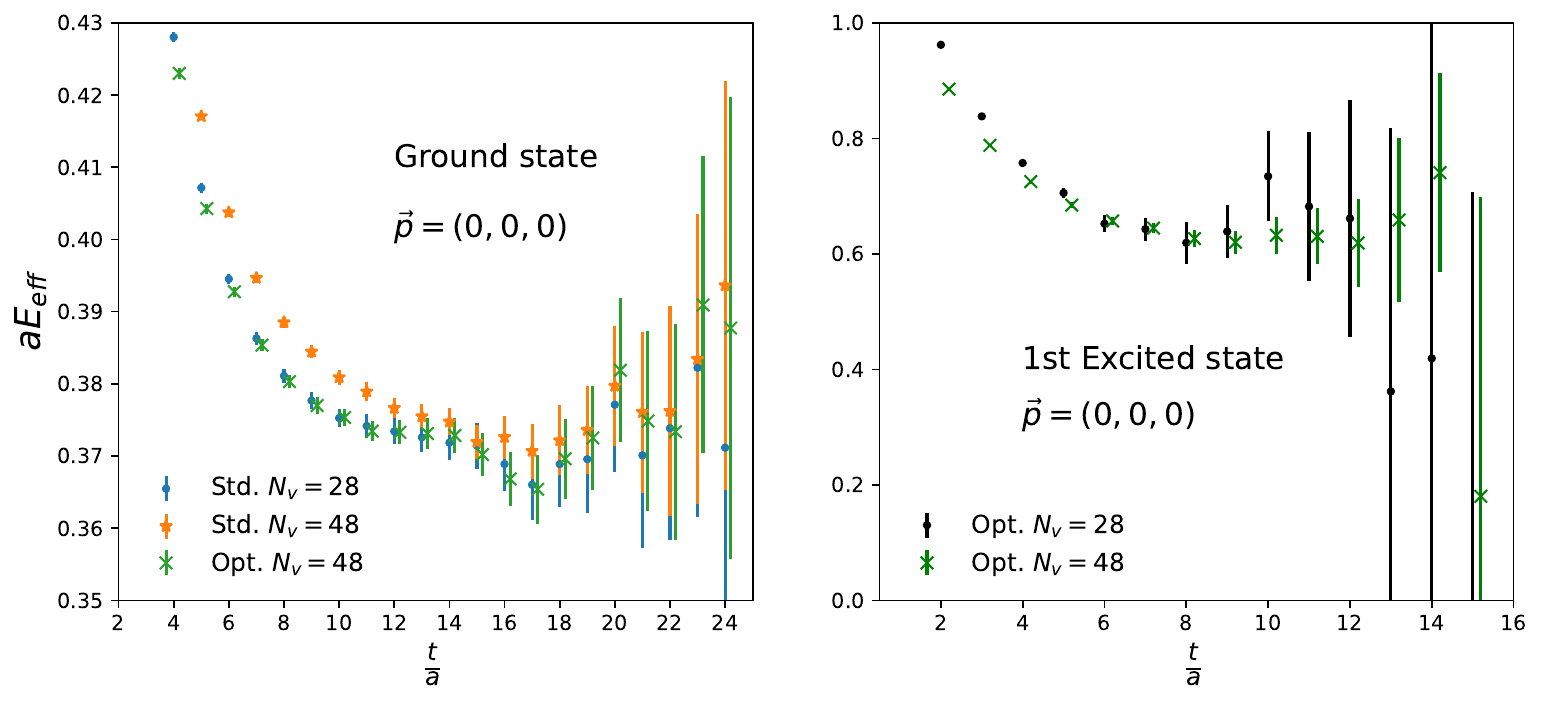}
    \caption{Effective masses of the ground and first excited states of the $\vec{p}=\vec{0}$ $G_1^{+}$ obtained with standard distillation (Std.) and optimal distillation profile (Opt.) at different choices of $N_v$.}
    \label{fig:NucleonMasses000}
\end{figure}

\begin{figure}
    \centering
    \includegraphics[width=0.9\textwidth]{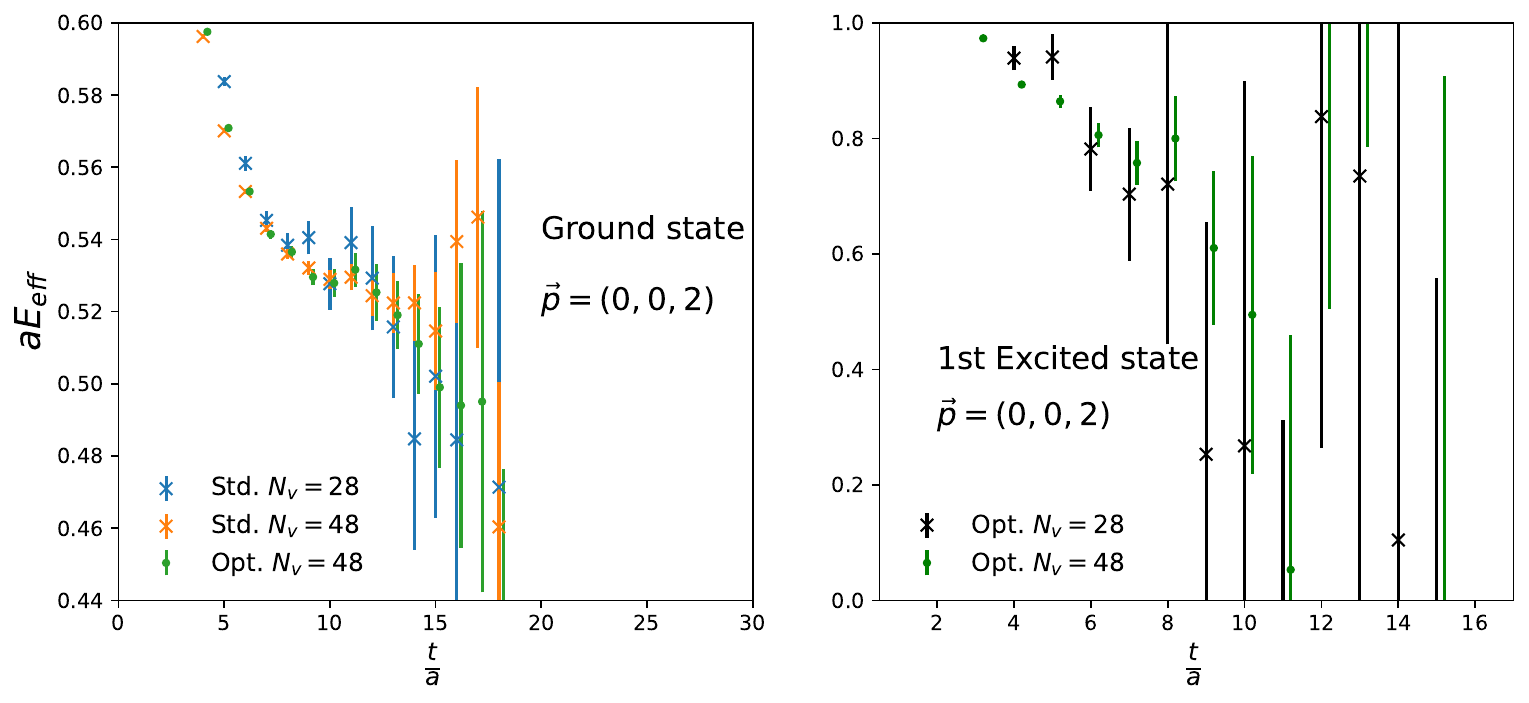}
    \caption{Effective masses of the ground and first excited states of the $\vec{p}=(0,0,2)$ $G_1$ obtained with standard distillation (Std.) and optimal distillation profile (Opt.) at different choices of $N_v$.}
    \label{fig:NucleonMasses001}
\end{figure}

State-of-the-art single- and multi-nucleon spectroscopy and scattering calculations often rely on distillation to build creation operators with large overlap onto the finite-volume energy eigenstates of interest, see e.g. \cite{BaryonScattering:2025ziz, Zhang:2026cco}. Here we extend the optimal distillation profile framework \cite{Knechtli:2022bji}, which has been successfully used by us for light mesons and charmonium \cite{Urrea-Nino:2025afu, Urrea-Nino:2026cjj} among other hadronic systems, to the case of nucleons. We build a basis of $N_B$ different nucleon elementals (or triplets) $\Phi^{(a)}(\vec{p},t)$ ($a=1,...,N_B$) introducing different quark distillation profiles to the original nucleon elementals $\Phi(\vec{p},t)$, i.e.
\begin{align*}
    \Phi[\vec{p},t]_{ijk} \rightarrow \Phi[\vec{p},t]^{(a)}_{ijk}=  g_i^{(a)}[t] \cdot  g_j^{(a)}[t] \cdot  g_k^{(a)}[t] \cdot \Phi[\vec{p},t]_{ijk},
\end{align*}
where $g_i^{(a)}[t]$ is the $a$-th quark profile evaluated at distillation index $i$ and time $t$, and $\vec{p}$ is the spatial momentum. There is no sum over $a$, $i$, $j$ and $k$. We omit any spin indices in the elementals since these are not affected by the distillation profiles. These profiles can be a function of the 3D Laplacian eigenvalues \cite{Knechtli:2022bji} or of the distillation index \cite{Urrea-Nino:2025ijv}. With these different elementals we compute an $N_B \times N_B$ correlation matrix and solve a GEVP to extract the low-lying energies of interest. This modification comes at very little additional computational cost, allows an optimization for each nucleon operator and state of interest, and can be extended to multi-hadron operators including one or more nucleons or baryons in general. We test this method on a subset of configurations from the CLS ensemble $X252$, also used in \cite{Alharazin:2026lno}, using $N_v = 48$ distillation vectors. Fig.~\ref{fig:NucleonMasses000} shows effective masses for the $G_1^{+}$ ground and first excited states at $|\vec{p}|=0$. For the ground state, we compare the masses using standard distillation (Std. $N_v=48$) to those using the optimal profile (Opt. $N_v=48$). The optimal profile leads to a significantly faster convergence to a plateau region. We also include the masses from standard distillation using $N_v=28$, which converge as fast as the case with the optimal profile. Clearly reducing $N_v$ here already has a positive effect for the ground state, however there is a caveat. Reducing $N_v$ proves detrimental for the first excited state, which we extract with an optimal profile GEVP for both $N_v=28,48$. Clearly, using $N_v = 48$ outperforms $N_v = 28$ in terms of statistical errors and excited-state contamination. The optimal profiles guarantee all available vectors are optimally used for each channel and excitation. Fig. \ref{fig:NucleonMasses001} shows the same results for $\vec{p}=(0,0,2)$; a similar pattern of improvement is seen as for the $\vec{p}=\vec{0}$ case. At larger values of momenta, the suppression of excited-state contamination starts to saturate because larger values of $N_v$ are required. Similar improvements were reported by J. Crawford \textit{et al.} at this conference, who utilize distillation profiles for a variational approach to nucleon matrix elements.

\subsection{Improved glueball operators}
\begin{figure}
    \centering
    \begin{subfigure}{0.495\textwidth}
        \centering
        \includegraphics[width=\linewidth,page=1]{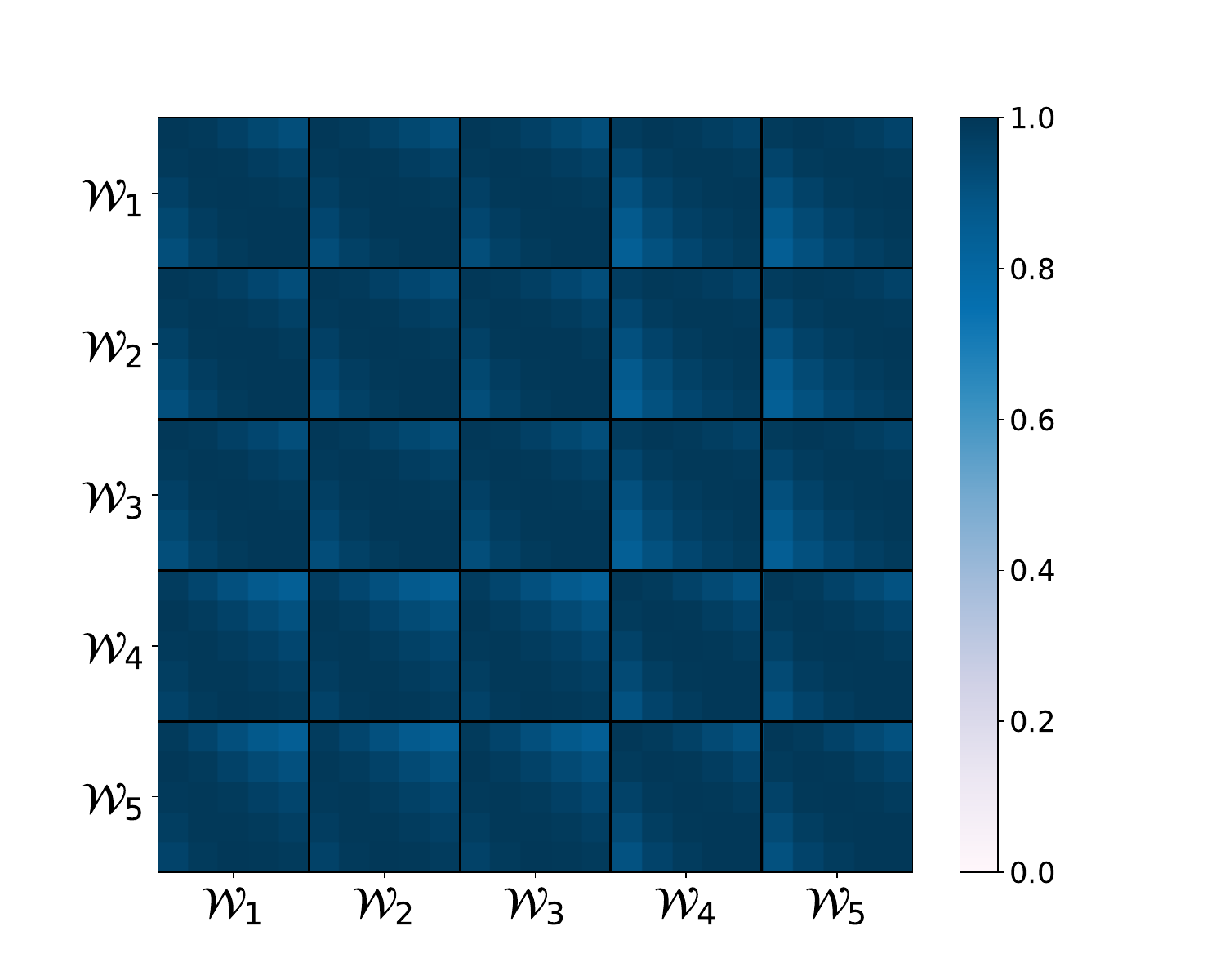}
    \end{subfigure}%
    \hspace{0pt}%
    \begin{subfigure}{0.495\textwidth}
        \centering
        \includegraphics[width=\linewidth,page=1]{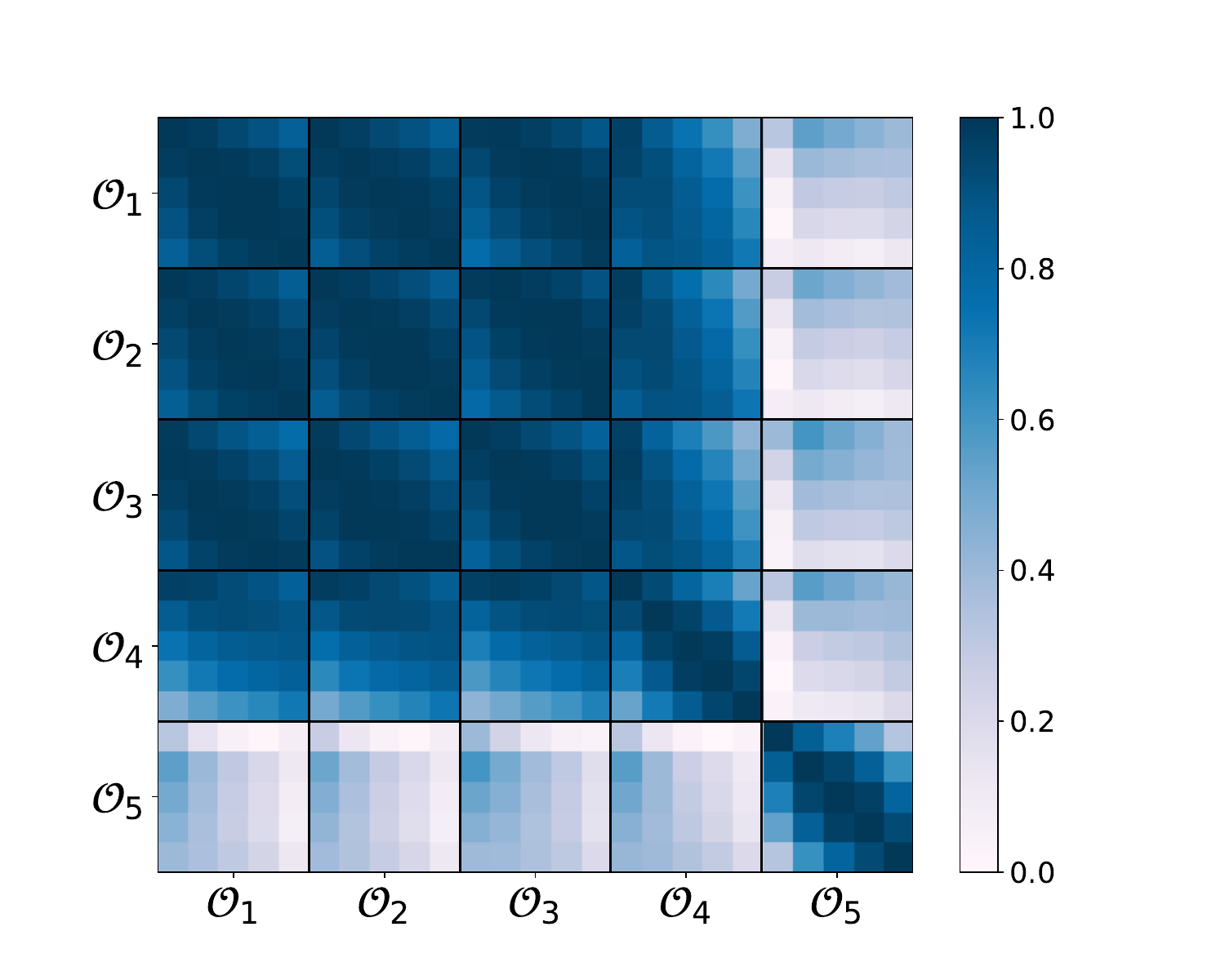}
    \end{subfigure}
    \caption{Absolute value of the correlation matrices at $t=a$ for 5 loop shapes under 5 levels of smearing (left) and 5 operators based on $\mathcal{B}_i$ and $D_i$ under 5 levels of smearing (right) for the $A_1^{++}$.}
    \label{fig:CorrGlue}
\end{figure}

Glueball calculations in pure-gauge, e.g.\ \cite{Sakai:2022zdc}, and with dynamical quarks, e.g.\ \cite{Jiang:2022ffl}, include hadron creation operators made entirely of gauge link variables to build glueball-like states. The most common construction is based on spatial loops with different shapes chosen to generate an irreducible representation (irrep) of the cubic group when at rest or the corresponding little group for non-zero spatial momentum, together with some form of gauge link smearing. This approach allows building a wide basis of operators by considering several different shapes and smearing levels, but it turns out that smearing can lead to different shapes having close to degenerate correlators, resulting in ill-conditioned correlation matrices \cite{Sakai:2022zdc}. Furthermore, contrary to meson operator construction following the intuitive $\bar{q} \Gamma q$ form, the loop shapes do not provide an intuition on physical features of the states of interest nor do they carry any information on continuum total angular momentum as can be done for meson operators \cite{Dudek:2010wm}. In \cite{Urrea-Nino:2026cjj} we use operators built from the chromo-magnetic component $\mathcal{B}_i = \epsilon_{ijk} F_{jk}$ of the gluon field-strength tensor and the adjoint spatial gauge-covariant derivative coupled to have a definite total angular momentum $J$ under SO(3) and then subduced onto the corresponding lattice irreps \cite{Chen:2005qb}. Instead of representing the operators (up to a fixed power of the lattice spacing $a$) on the lattice via linear combinations of different loop shapes \cite{Liu:2001je}, we discretize the field-strength tensor via the clover definition and the derivatives via gauge-covariant symmetric differences \cite{Lee:2002fj}. This makes the implementation more straightforward, since symmetric differences require only nearest-neighbor relations while the linear combination of loops might require further-separated points, making a parallel code significantly harder to write~\cite{Morningstar:1999rf}. Furthermore, since these operators preserve information from the continuum, contain different amounts of fields and derivatives to sample different spatial structures, and don't require projection coefficients for different irreps, they are considerably more advantageous than spatial loops both in terms of physics and implementation, particularly for studies which necessitate large statistics. In \cite{Urrea-Nino:2026cjj} we list our choice of operators for $A_1^{++}$. For $A_1^{-+}$ we have generated five operators, beyond the single one listed in \cite{Chen:2005qb}, using the \textit{opBasis} library \cite{Husung:2025gfk}, which are given by
\begin{align*}
\begin{aligned}
    \mathcal{O}_1(t) &= \sum_{\vec{x}}\text{Tr}\left( \epsilon_{ijk} \mathcal{B}_i(\vec{x},t) D_j \mathcal{B}_k(\vec{x},t)   \right)\\
    \mathcal{O}_2(t) &= \sum_{\vec{x}} \text{Tr} \left( \epsilon_{ijk} \epsilon_{ilm}\mathcal{B}_j(\vec{x},t) \mathcal{B}_k(\vec{x},t)  D_l \mathcal{B}_m(\vec{x},t) \right)\\
    \mathcal{O}_3(t) &= \sum_{\vec{x}} \text{Tr}\left( \epsilon_{ijk} \mathcal{B}_i(\vec{x},t) D_j^3 \mathcal{B}_k(\vec{x},t)  \right)
\end{aligned}
\qquad
\begin{aligned}
    \mathcal{O}_4(t) &= \sum_{\vec{x}} \text{Tr}\left( \epsilon_{ijk} \mathcal{B}_i(\vec{x},t) D_j^2 D_k \mathcal{B}_j(\vec{x},t)   \right)\\
    \mathcal{O}_5(t) &= \sum_{\vec{x}} \text{Tr}\left( \epsilon_{ijk} \mathcal{B}_i(\vec{x},t) D_j D_i^2 \mathcal{B}_k(\vec{x},t)  \right),
\end{aligned}
\end{align*}
where repeated indices are summed over. As for the $A_1^{++}$ case, these independent operators are built by fixing a mass-dimension and finding all possible different nonequivalent couplings of $\mathcal{B}_i$ and $D_i$ with the correct quantum numbers. In Fig. \ref{fig:CorrGlue} we show the correlation function at $t=a$ for both 5 different loop shapes ($\mathcal{W}_i$) and 5 different  $A_1^{++}$ operators used in \cite{Urrea-Nino:2026cjj} ($\mathcal{O}_i$), in both cases measuring the operators at 5 different levels of spatial APE smearing. $\mathcal{O}_5$ remains substantially decoupled from the other four, regardless of the amount of smearing, thereby overcoming the near-degeneracy problem faced by the conventional spatial loops. One reason is this operator couples more to a $J=4$ than to a $J=0$ state in the continuum.  Results for the $A_1^{-+}$ will be presented in a coming publication. 
\subsection{Improved charmonium operators}
\begin{figure}
    \centering
    \includegraphics[width=0.75\linewidth]{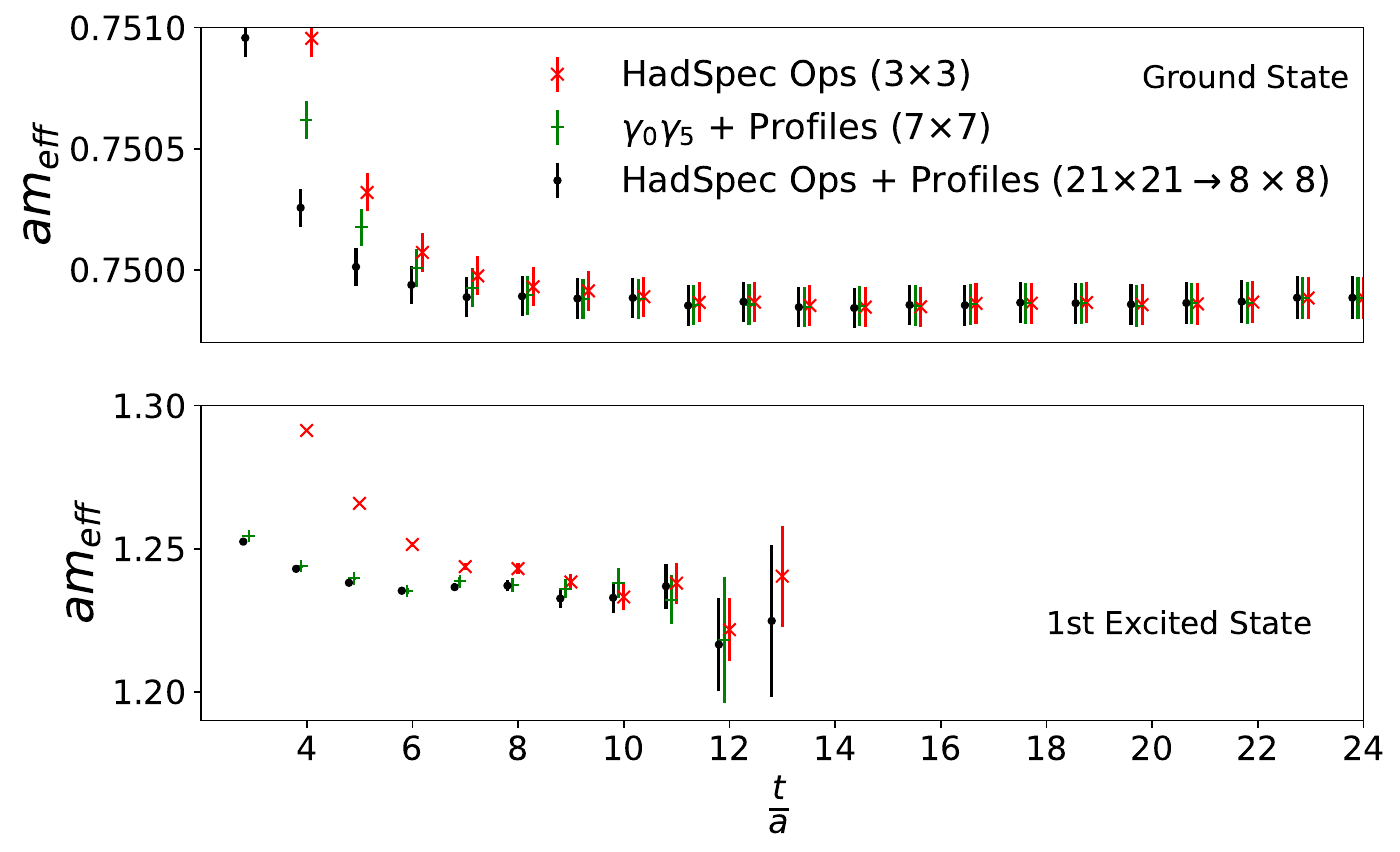}
    \caption{Effective masses of the ground and first excited states for the $I=1$, $A_1^{-+}$ charmonium channel using different GEVPs based on operators of the form $\bar{q}\Gamma q$. Red crosses $(\times)$: GEVP using three different $\Gamma$ with standard distillation, as done by the HadSpec collaboration. Green plus signs ($+$): $\Gamma =\gamma_0 \gamma_5$ using 7 different Gaussian distillation profiles. Black dots ($\cdot$): three $\Gamma$ operators, each one with 7 Gaussian distillation profiles. In the last case, we prune the matrix from size $21\times 21$ to $8\times 8$ as described in the text before solving the GEVP.}
    \label{fig:OneHadron}
\end{figure}

\begin{figure}
    \centering
    \includegraphics[width=0.75\linewidth]{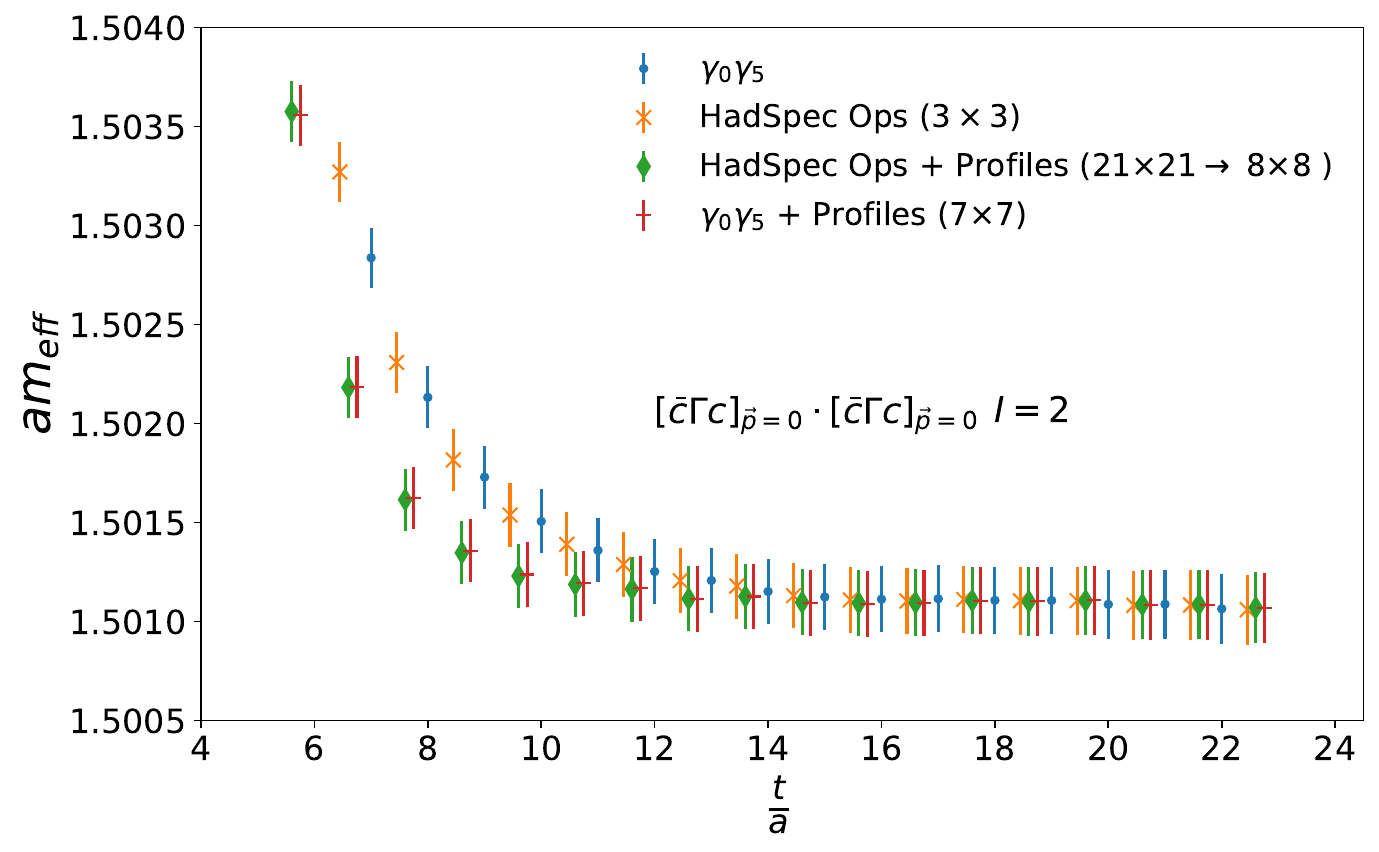}
    \caption{Ground-state effective masses of the $I=2$, $A_1^{++}$ channel from different two-meson operators at zero total and back-to-back momenta built from one-meson operators. Blue dots use $\Gamma = \gamma_0 \gamma_5$, orange crosses use the optimal one-meson operator involving $\Gamma = \gamma_0 \gamma_5, \gamma_0 \gamma_5 \gamma_i \nabla_i, \gamma_i \mathbb{B}_i$ with standard distillation, the red plus signs use the optimal one-meson operator involving $\Gamma = \gamma_0 \gamma_5$ and Gaussian distillation profiles, and the green diamonds use the one involving $\Gamma = \gamma_0 \gamma_5, \gamma_0 \gamma_5 \gamma_i \nabla_i, \gamma_i \mathbb{B}_i$ with Gaussian distillation profiles.}
    \label{fig:I2Masses}
\end{figure}

The ongoing multi-hadron scattering calculations have emphasized again the importance of using operators which lead to reliable mass plateaus and avoid missing intermediate states in the spectrum \cite{Green:2026tf}. Since the method of optimal distillation profiles has been widely successful in single-hadron calculations, we extend it here for the first time to the case of two-meson operators in $N_f=2$ QCD with $m_\pi \approx 2.2$ GeV in a lattice of size $24^3 \times 48$, $a\approx 0.066$ fm and using $N_v = 200$ \cite{Knechtli:2022bji}, improving on a set of derivative-based operators widely used by the \textsc{Hadron Spectrum Collaboration} \cite{Dudek:2010wm}. Following the strategy of \cite{Wilson:2023anv}, we first build optimal single-meson operators $\bar{q}\Gamma q$ which are then put together into two-meson operators. Fig.\ \ref{fig:OneHadron} shows the ground and first excited state effective masses for the isovector ($I=1$) $A_1^{-+}$ charmonium obtained via three different GEVP approaches. The red crosses $(\times)$ include standard distillation with three operators; $\Gamma = \gamma_0 \gamma_5, \gamma_0 \gamma_5 \gamma_i \nabla_i, \gamma_i \mathbb{B}_i$, with $\nabla_i$ the spatial gauge-covariant derivative in direction $i$ and $\mathbb{B}_i = \epsilon_{ijk} \nabla_j \nabla_k$, see the operator basis used in \cite{Wilson:2023anv}. The green plus signs ($+$) include $\gamma_0 \gamma_5$ with $N_B = 7$ different Gaussian distillation profiles. The black dots ($\cdot$) include the three $\Gamma$, each one with $N_B=7$ different Gaussian profiles, resulting in a $21\times 21$ correlation matrix. We prune this matrix down to an $8\times 8$ one using the singular vectors corresponding to the $8$ largest singular values extracted at a fixed value of time separation \cite{Balog:1999ww} before solving the GEVP. For both resolved states, the use of profiles with a single $\Gamma$ outperforms using multiple $\Gamma$s with standard distillation in terms of faster convergence to the plateau. The improvement is particularly large for the first excited state. We use the optimal operators built from each GEVP to build an $I=2$, $A_1^{++}$ two-meson operator with zero total and back-to-back spatial momentum. Since we do not solve a GEVP with the two-meson operators, we cannot expect optimality in convergence to a plateau. However, we do expect our optimized single-meson operators to perform better than the standard ones when building the two-meson operators. In Fig.\ \ref{fig:I2Masses} we show the effective masses coming from a single two-meson operator built from the different optimized single-meson operators as defined by the previously mentioned GEVPs. The blue dots $(\cdot)$ use $\Gamma = \gamma_0 \gamma_5$ with standard distillation, the orange crosses $(\times)$ use the optimal operator using $\Gamma = \gamma_0\gamma_5, \gamma_0 \gamma_5 \gamma_i \nabla_i, \gamma_i \mathbb{B}_i$ with standard distillation, the red plus signs $(+)$ use $\Gamma = \gamma_0 \gamma_5$ with an optimal profile and the green diamonds $(\blacklozenge)$ use the optimal operator from $\Gamma = \gamma_0\gamma_5, \gamma_0 \gamma_5 \gamma_i \nabla_i, \gamma_i \mathbb{B}_i$ and the Gaussian profiles for each one. The use of profiles, either with one or multiple $\Gamma$, yields a significantly faster convergence to the plateau, while for standard distillation there is no major difference between using one or three $\Gamma$s. The improvement brought forth by using profiles, even when only using a single $\Gamma$, gives a promising perspective for optimizing multi-meson calculations beyond the case studied in this work. As shown by J. Koponen \textit{et al.} at this conference, using optimal distillation profiles can substantially reduce excited-state contamination for charmonium operators with $|\vec{p}|^2 > 0$, which further motivates their use for multi-meson scattering calculations. 

\section{Conclusions}
We presented here significant improvements in operator construction for three hadronic systems of interest for modern lattice QCD spectroscopy calculations: nucleons, glueballs and multi-meson systems. First, the use of optimal distillation profiles resulted in a major reduction of excited-state contamination in nucleon correlation functions at different spatial momenta, allowing to access physically useful information at short time separations to avoid the signal-to-noise problem at large time separations. Second, we used and extended a set of glueball-like operators built to preserve total angular information from the continuum, as well as presented a novel and efficient implementation for them. The coupling of chromo-magnetic fields and adjoint derivatives allowed to overcome the problem of near degeneracies in correlation matrices when using conventional spatial loops with spatial link smearing. Third, charmonium operators with optimal distillation profiles were used for the first time to build improved two-meson operators whose effective masses displayed a much faster convergence to the plateau compared to operators built using standard distillation together with derivative-based constructions. These improvements exploit additional physics-based degrees of freedom, such as distillation profiles for nucleons and mesons as well as subduction of continuum angular momentum for glueballs. Finally, they are realized at very little computational cost and with efficient implementations, making them ideally suited for large-scale spectroscopy and scattering studies where reliable energy determinations are a fundamental first step. 
\par
\section*{Author contribution statement}
J.A.U.-N. contributed with theoretical operator construction, code development, measurement of correlations, data analysis and preparation of results to all three sections of this work. F.K., T.K, and M.P. contributed to the work on glueball and multi-meson calculations. N.H. contributed to the work on glueball operators. F.R.-L., J.B. and M.S. contributed to the work on nucleons. 
\section*{Acknowledgements}
J.A.U.-N. acknowledges support from a Research Ireland (Science Foundation Ireland) Frontiers for the Future Project award [grant number SFI-21/FFP-P/10186]. 
J.B. is supported by the European Research Council (ERC) consolidator grant StrangeScatt-101088506. The work of F.R.L. and M.S. was supported in part by the Swiss National Science Foundation (SNSF) through grant No. 200021-236432. N.H. acknowledges support by the projects PID2021-127526NB-I00, funded by MCIN/AEI/10.13039/501100011033 and by FEDER
EU, as well as IFT Centro de Excelencia Severo Ochoa No CEX2020-001007-S, funded by
MCIN/AEI/10.13039/501100011033. Part of the work on glueball and multi-meson operators was supported by the German Research Foundation
(DFG) research unit FOR5269 "Future methods for studying confined gluons in QCD". J.A.U.-N., F.K. and T.K. gratefully acknowledge the Gauss Centre for Supercomputing e.V. (www.gausscentre.eu) for funding this project by providing computing time on the GCS Supercomputer SuperMUC-NG at Leibniz Supercomputing Centre (www.lrz.de) under GCS/LS project ID pn29se for computing the distillation vectors, perambulators, elementals and correlation functions for the multi-meson study, as well as computing time and storage on the GCS Supercomputer JUWELS at Jülich Supercomputing Centre (JSC) under GCS/NIC project IDs HWU35 and HWU17 for this same study. Correlation functions for the nucleons were calculated using a grant from the Swiss National Supercomputing Centre (CSCS) under project ID lp53 on Alps. Perambulators for the nucleons were computed on UBELIX (https://www.id.unibe.ch/hpc), the HPC cluster at the University of Bern, using the QUDA-LapH package \cite{CosmonCollaboration2025quda_laph}. Elementals/triplets for the nucleons were computed on the HPC cluster Elysium of the Ruhr-Universität Bochum, subsidized by the DFG (INST 213/1055-1). We thank colleagues in the CLS collaboration for sharing the ensemble X252 used for the nucleon calculations. We also thank Joshua Crawford and members of the BaSc collaboration, in particular Jeremy Green, for useful discussions about nucleons with distillation profiles.

\bibliographystyle{JHEP_arXiv}
\bibliography{refs}

\end{document}